\documentclass[aps,twocolumn,prd,superscriptaddress,letterpaper]{revtex4-1}
\usepackage{amsmath, amssymb}
\usepackage{amsfonts}
\usepackage{amsmath}
\usepackage{mathrsfs}
\usepackage[usenames,dvipsnames]{xcolor}     
\usepackage{parskip}
\usepackage{graphicx,float}
\usepackage[german, english]{babel}
\usepackage{xspace}
\usepackage{tabularx}
\usepackage{textcomp}
\usepackage{url}
\usepackage{bm}
\usepackage[pdftex,  pagebackref=true]{hyperref}
\usepackage{graphicx}
\usepackage{subfigure}
\usepackage{listings}
\usepackage[dvipsnames]{xcolor}

 \definecolor{linkcolor}{rgb}{.8,0,0}
 \definecolor{urlcolor}{rgb}{0,0,.7}
 \definecolor{citecolor}{rgb}{0,.5,0}
 \definecolor{acrocolor}{rgb}{0,0,.7}

 \hypersetup{bookmarksopen,colorlinks=true}
 \hypersetup{pdfstartview=FitH}
 \hypersetup{linktocpage=true,bookmarksnumbered=true}
 \hypersetup{plainpages=false,breaklinks=true}
 \hypersetup{linkcolor=linkcolor,citecolor=citecolor,urlcolor=urlcolor}

\begin{document}

\title{Simplified optical configuration for a sloshing-speedmeter-enhanced gravitational wave detector}

\author{Andreas Freise}\email{andreas.freise@ligo.org}
\author{Haixing Miao}
\affiliation{School of Physics and Astronomy and Institute for Gravitational Wave Astronomy, University of Birmingham, Edgbaston, Birmingham B15 2TT, United Kingdom}
\author{Daniel D. Brown}
\affiliation{Department of Physics and The Institute of Photonics and
  Advanced Sensing (IPAS) University of Adelaide, SA, 5005, Australia,
  and OzGrav, Australian Research Council Centre of Excellence for Gravitational Wave Discovery}

\date{\today}

\begin{abstract}
\noindent
We propose a new optical configuration for an interferometric
gravitational wave detector based on the speedmeter concept using a 
sloshing cavity.  Speedmeters provide an inherently
better quantum-noise-limited sensitivity at low frequencies than the
currently used Michelson interferometers. 
We show that a practical sloshing cavity can be added 
relatively simply to an existing dual-recycled Michelson
interferometer such as Advanced LIGO.
\end{abstract}

\maketitle

Current interferometric gravitational wave detectors such as LIGO, 
Virgo, GEO or KAGRA~\cite{AdvancedLIGO15, AdvancedVirgo15, Grote2010,
  Aso13} are based on modified Michelson interferometers. The design
for future gravitational wave observatories such as the Einstein
Telescope~\cite{Punturo10} requires the use of more advanced 
quantum-noise-reduction techniques. 

The Michelson interferometer is sensitive to the difference in
position of the end test masses. The concept of the speedmeter
refers to interferometer configurations whose main output signal
is proportional to the speed of the end test masses instead.
Speedmeters by themselves cannot provide quantum-noise reduction 
at the desired level. Both position- and  speedmeters can be improved
by additional optical systems, such as for example squeezed light
and filter cavities. However, the speedmeters provide an inherently
better quantum-noise-limited sensitivity at low frequencies, and
thus might lead to less complex designs for future detectors.
In this paper we discuss a
new variant of the speedmeter concept and investigate practical
issues of this configuration.
Based on the idea of using a so-called sloshing cavity to turn a
Michelson interferometer into a speedmeter, we investigate 
alternative optical configurations for coupling the sloshing
cavity to the main interferometer. In particular we show 
a new configuration that is equivalent to previously proposed
in terms of quantum-noise reduction, but features fewer
additional optics. We briefly discuss the length noise
requirements for the sloshing cavity.

\section[]{Sloshing Michelson as a Speedmeter}

\begin{figure}[htb]
	\begin{center}
		\includegraphics[width=0.485\textwidth ]{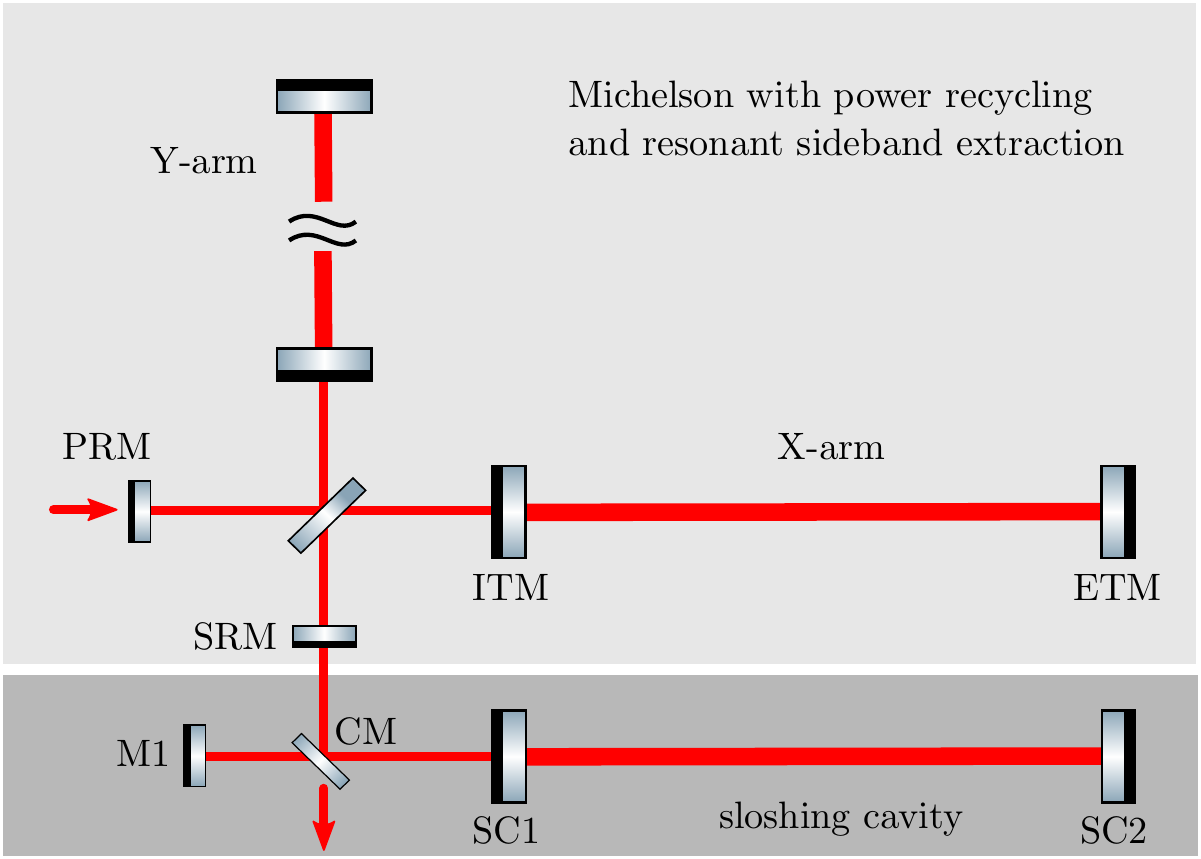}
	\end{center}
	\caption{Optical layout of one possible Advanced LIGO design with a sloshing cavity as described in
  ~\cite{Purdue02} (sloshing RSE MI).}
	\label{fig:layout_DRMI}
\end{figure}

The concept of using a so-called sloshing cavity with a Michelson
interferometer to generate a speedmeter was first introduced by
Purdue and Chen~\cite{Purdue02}. The optical layout of one
version of this scheme is shown in Figure~\ref{fig:layout_DRMI}.
The Advanced LIGO layout is retained, using a  Signal Recycling 
Mirror (SRM) with different transmissivity, see below, and the
output light is coupled to a so-called sloshing cavity. 
The coupling beam splitter (CM) allows some light
to reach the output, while some light is directed back and forth
between the interferometer and the sloshing cavity. Another mirror
(M1) is used to close the other open port. 

With the correct tuning of transmissions and postion of the new
optics, the sloshing cavity returns the signal light back into the 
interferometer, such that the relative mirror positions are measured
again at a different time, and the light leaking out from CM
to the output produces  a speedmeter signal, see \cite{Purdue02}
for a detailed description. 

In the following we will refer to this setup as \textit{original sloshing RSE
  MI} (sloshing resonant-sideband extraction Michelson).
In this setup the SRM transmissivity has
been set equal to that of the Input Test Masses (ITMs) of
the interferometer arms, and they form an impedance matched
cavity. This modifies the interferometer response such that the signal
light does not experience a frequency dependent change due to 
the signal recycling cavity or the arm cavity, i.e. $T_{\rm SRC}=1$. 
This choice provides intuitive mathematical descriptions but does
not represent an optimal optical design, as we will show below.
The original proposal should be understood as a conceptual design that 
introduces the idea rather than a practical design. For example, 
the longitudinal and alignment degrees of freedom of
each additional mirror must be controlled by feedback systems to
very high precision (as with all main optical elements of the Michelson
interferometer). A simple optical layout with fewer optics
will significantly help with designing feedback systems without introducing 
additional disturbances or unwanted coupling between optics. 

\begin{table}[htb]
\centering
\begin{tabular}{l|p{13mm}|p{17mm}|p{17mm}|p{17mm}}
\hline
\hline
parameter & aLIGO reference & orig. sloshing DRMI & sloshing alternative 1& sloshing alternative 2\\
\hline
power in& 125\,W & 125\,W & 250\,W & 125\,W\\
arm power & 770\,kW & 770\,kW & 770\,kW& 770\,kW\\
BS power& 5.5\,kW & 5.5\,kW & 100\,kW & 5.5\,kW\\
$L_{\rm arm}$ & 3995\,m & 3995\,m & 3995\,m & 3995\,m\\
$T_{\rm PRM} $& 0.03 & 0.03 & 0.0072 & 0.03\\
$T_{\rm ITM}$ & 0.014 & 0.014& 0.22 & 0.014\\
$T_{\rm ETM}$ & 5\,ppm & 5\,ppm & 5\,ppm& 5\,ppm\\
$F_{\rm arm}$ & 450 & 450 & 24 & 450\\
$T_{\rm SRM}$ & 0.2 & 0.014 & 1 & 0.2\\
$T_{\rm SRC}$ & 0.22 & 1 & 1 & 0.22\\
$\Omega$ & - & $2\pi\cdot 200$\, Hz& $2\pi\cdot 200$\, Hz& $2\pi\cdot 200$\,Hz\\
$\delta $& - & $2\pi\cdot 750$\, Hz & $2\pi\cdot 750$\, Hz& $2\pi\cdot 750$\,Hz\\
$T_{\rm CM} $& - & 0.063 & 0.996 & 0.996\\
$T_{\rm SC1} $& - & 0.0011 & 1 & 1\\
$T_{\rm SC2} $& - & 0 & 0 & 0 \\
$F_{\rm SC}$ & - & 5600 & 5600 & 5600\\
\hline
\hline
\end{tabular}
\caption{Optical parameters of the example setups discussed in this
  paper: a) the Advanced LIGO reference design and b)
  the original sloshing cavity setup as shown in
  Figure~\ref{fig:layout_DRMI}. In addition the parameters for
two new alternative configurations are shown, see the main text for a
detailed description.}
\label{tab:params_DRMI} 
\end{table}

For comparing the quantum-noise performance of the interferometer
it is useful to define the bandwidth  of the interferometer
by the transmissivities of  the coupling mirror (CM) and the 
sloshing cavity input mirror (SC1)
 (see also
Appendix~\ref{sec:bandwidth}):
\begin{equation}
\Omega = \frac{c\,\sqrt{T_{\rm SC1}}}{2 L}
\end{equation}
and
\begin{equation}
\delta = c\, T_{\rm CM}/L
\end{equation}
with $L$ the length of the interferometer arms, which is also the
length of the sloshing cavity.

In practice we have a desired or required values for $\delta$ and $\Omega$
from which we can compute the trasmissivities of
the sloshing cavity optics as:
\begin{equation}
T_{\rm CM}=\delta \, L / c \qquad \mbox{and} \qquad T_{\rm SC1} = (2\Omega\,L/c)^2
\end{equation}

\begin{figure}[htb]
	\begin{center}
		\includegraphics[width=0.485\textwidth ]{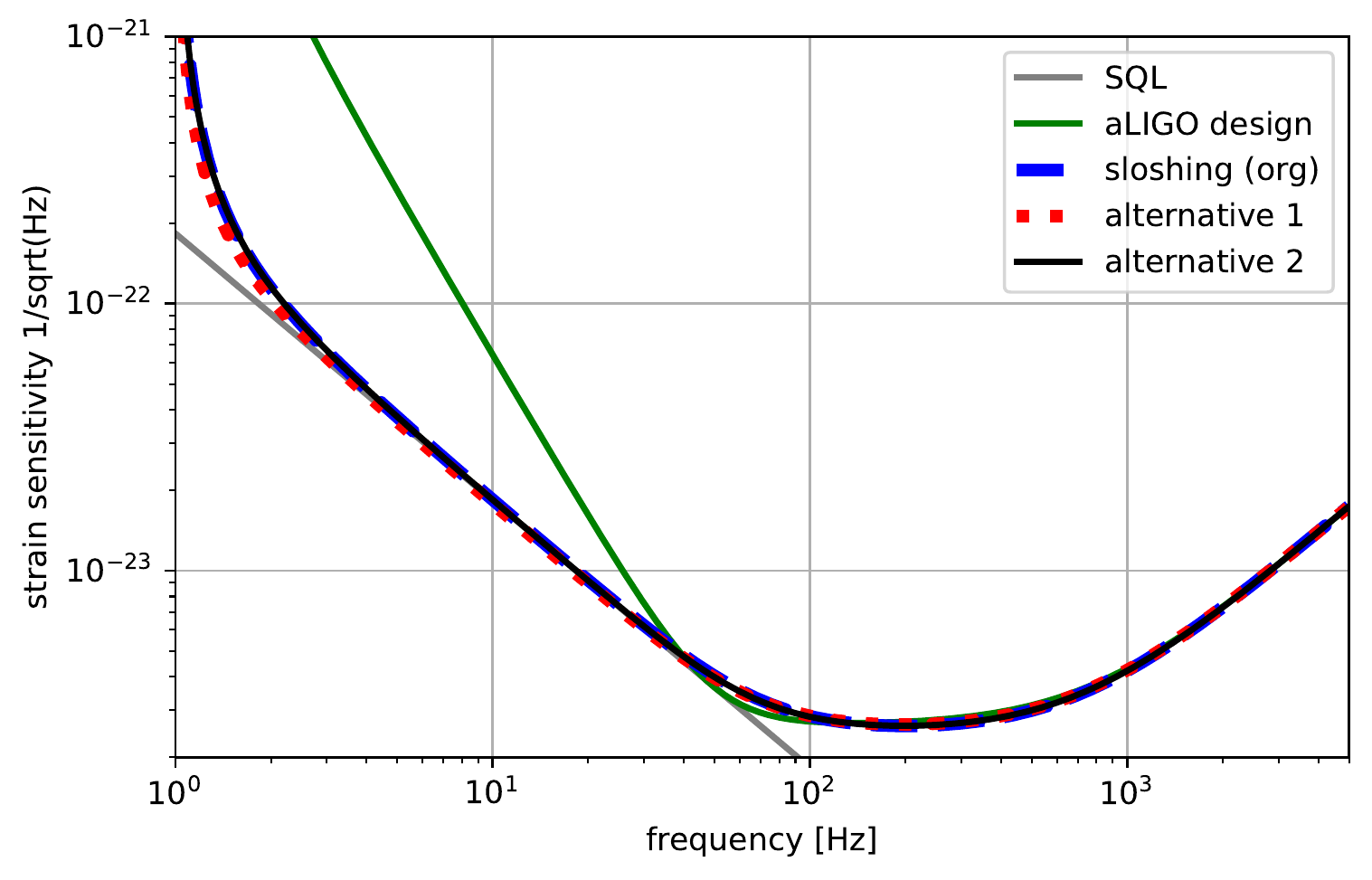}
	\end{center}
	\caption{Quantum-limited sensitivity of the Advanced LIGO design
    with different sloshing cavity designs, the
    optical parameters are listed in Table~\ref{tab:params_DRMI}. The 
    design sensitivity of Advanced LIGO (aLIGO) and the Standard
    Quantum Limit (SQL) are shown for reference.}
	\label{fig:sens_compare}
\end{figure}

For a lossless sloshing cavity we can generate an optimised result with
an equal sensitivity at high frequencies to the Advanced LIGO design
using the following parameters: $\delta = 2\pi\times 750$\, Hz and 
$\Omega = 2\pi\times 200$\,Hz.  The quantum-noise limited sensitivity
is shown in Figure~\ref{fig:sens_compare}, and the optical
parameters are listed in Table~\ref{tab:params_DRMI}. The sloshing
cavity in this example has the same length (3995\,m) as the
interferometer arm cavities, and has a finesse of $F=5600$.
The quantum-noise-limited sensitivity of the setup shows the 
typical shape of a speedmeter configuration with the quantum 
noise touching the Standard Quantum Limit (SQL) for a wide 
range of frequencies, significantly improved over the 
RSE Michelson.

\section{Alternative configurations}
We can quickly 
compare a wide range of optical layouts by
using numerical interferometer modelling software such as
\textsc{Finesse}~\cite{Freise04, Finesse}.
We used simple
analytical derivations for the interferometer bandwidth 
to find setups which should have an equivalent performance
and then confirmed this with a numerical model that does
not rely on our assumptions or approximations.

\begin{figure}[htb]
	\begin{center}
		\includegraphics[width=0.485\textwidth ]{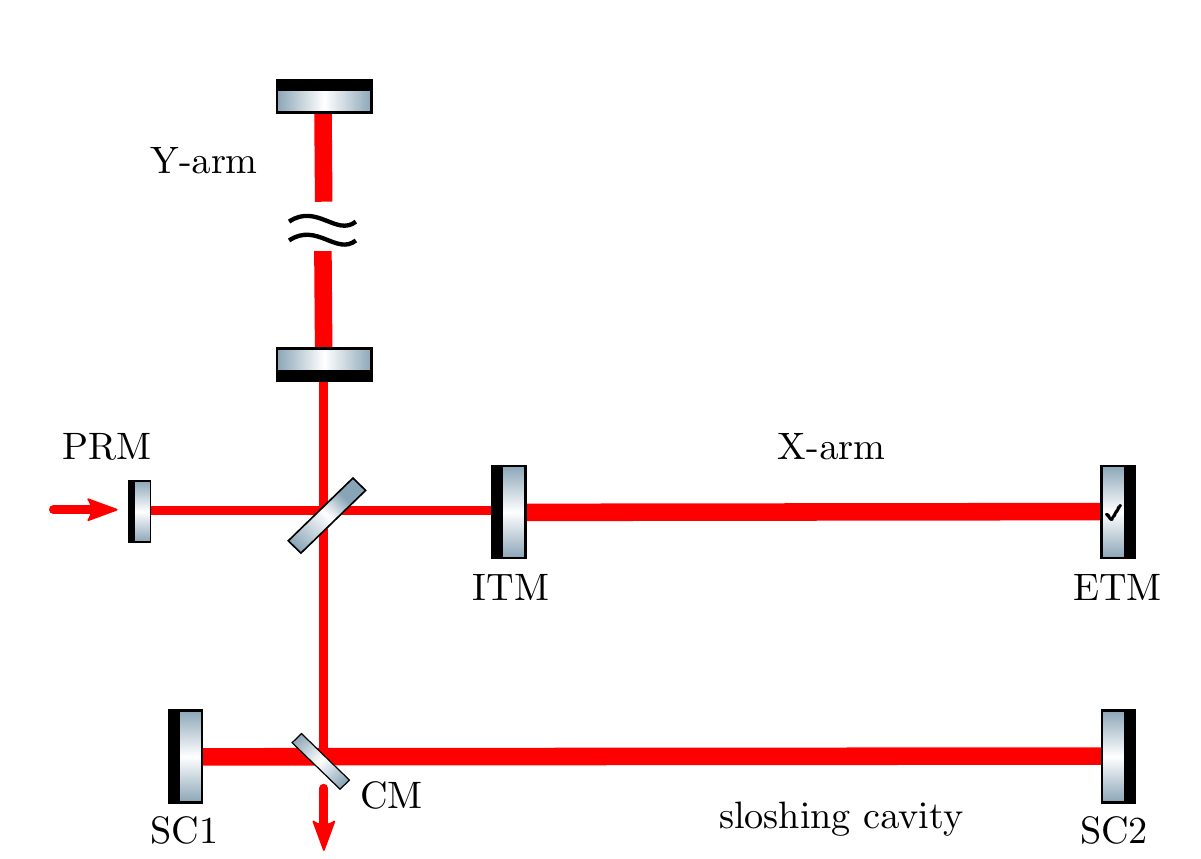}
	\end{center}
	\caption{Alternative optical layout for  sloshing cavity based on
   the Advanced LIGO design (sloshing MI).}
	\label{fig:layout_MI_optimised}
\end{figure}
\begin{figure}[htb]
	\begin{center}
		\includegraphics[width=0.485\textwidth ]{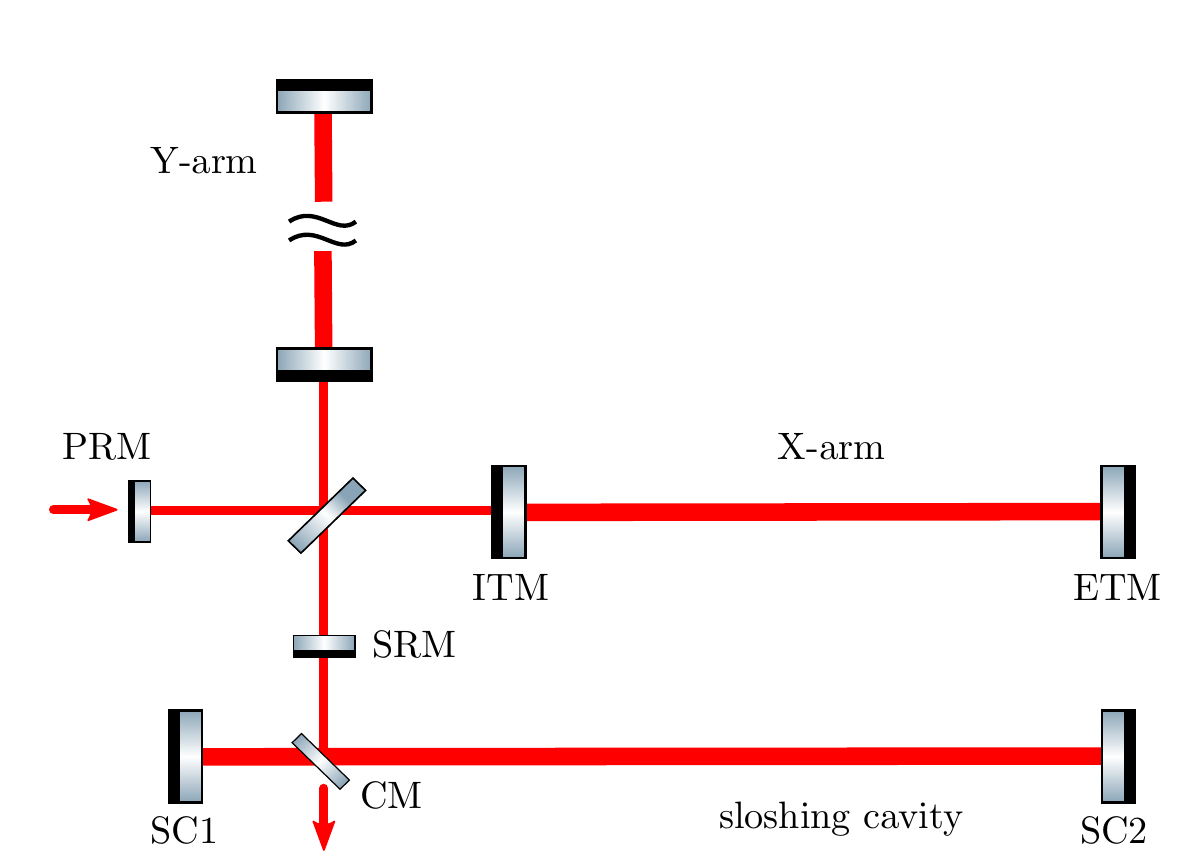}
	\end{center}
	\caption{Second alternative optical layout for a sloshing cavity based on
    the Advanced LIGO design, including a signal recycling mirror and
    a coupling mirror inside the sloshing cavity.}
	\label{fig:layout_MI_power_optimised}
\end{figure}

We can achieve the same quantum-noise-limited sensitivity with 
simplified setups shown in Figures~\ref{fig:layout_MI_optimised} 
and~\ref{fig:layout_MI_power_optimised}.
The layout shown in  Figure~\ref{fig:layout_MI_optimised} 
has two significant changes to the one shown before:
a) the Signal Recycling Mirror has been removed and b) the
coupling mirror is located inside the sloshing cavity. The immediate
advantage of this new setup is that in includes two fewer mirrors 
and presents a much less complex control problem. 
We can compute the sloshing frequency and extraction rate in this
setup using the reflectivity and transmissivity of the sloshing
cavity derived in Appendix~\ref{app:sloshing_inside},
and $\Omega$ is now defined as:

\begin{equation}\label{eq:cubic1}
\Omega = \frac{c\,\sqrt{R_{\rm CM} T_{\rm ITM}}}{2 L}
\end{equation}
The bandwith of the system, i.e. the extraction rate $\delta$ will
be defined by the combined transmissions though the sloshing 
cavity $T_{\rm SC}$ and the transmission of the input test masses 
$T_{\rm ITM}$.

The reflectivity of the sloshing cavity is (Appendix~\ref{app:sloshing_inside}):
\begin{equation}\label{eq:cubic2}
r_{\rm SC}= \frac{1-T_{\rm CM}}{1+T_{\rm CM}}
\end{equation}
And the reflectivity of the coupled cavity formed by the ITM and
the sloshing cavity is given as:
\begin{equation}\label{eq:cubic3}
R_{\rm CC}= \frac{(r_{\rm ITM} - r_{\rm SC})^2}{(1-r_{\rm ITM}r_{\rm
    SC})^2}, \qquad
\end{equation}
The half bandwidth  of such a cavity is: 
\begin{equation}
\gamma = \frac{c}{2 L} \frac{1-r_{\rm
      CC}}{\sqrt{r_{\rm CC}}}
\end{equation}
For a given $\gamma=750\times 2 \pi$ this leads to a quadratic
equation for the reflectivity of the coupled cavity with only one real
solution: $R_{\rm CC}  = 0.778$.

With this result, $\Omega =200\times 2 \pi$, and from
equations~\ref{eq:cubic1}, ~\ref{eq:cubic2} and~\ref{eq:cubic3} we obtain 
a cubic equation with only one real solution, from which 
we obtain the values:
\begin{equation}
T_{\rm CM} = 0.995 \qquad\mbox{and}\qquad T_{\rm ITM} = 0.22
\end{equation}
And indeed, a numerical search for the optimal parameters, taking
into consideration the transmission of the ETMs and surface losses of
37\,ppm per surface in the main interferometer, provides very 
similar values:
\begin{equation}
T_{\rm CM} = 0.996 \qquad\mbox{and}\qquad T_{\rm ITM} = 0.228
\end{equation}
The resulting quantum-noise-limited sensitivity is shown in Figure~\ref{fig:sens_compare}.

The new layout has the advantage of using two  fewer mirrors but
requires a large circulating light power inside the power recycling
cavity and thus within the substrate of the main beam splitter.
Instead  we can achieve the same sensitivity if we use the same 
main interferometer parameters as the current Advanced LIGO layout,
i.e. a signal recycling mirror with moderate transmission and an arm cavity 
with a finesse of 450.
We can use of the fact that the bandwidth of the original
signal recycling cavity is already correct and that we have computed
the corresponding reflectivity of the coupling mirror already for
the first alternative layout. 

This optical layout is shown in 
Figure.~\ref{fig:layout_MI_power_optimised}, 
the resulting parameters for the mirror reflectivities are 
listed in Table~\ref{tab:params_DRMI}  as `sloshing alternative 2', 
and the corresponding
sensitivity is also shown in Figure.~\ref{fig:sens_compare}.
This second alternative
provides a more practical option: As shown in
Table~\ref{tab:params_DRMI} this configurations leaves the power
in the beam splitter unaltered whereas the first alternative requires
an almost 20 times higher power in the centreal beam splitter.
Experience at the GEO detector has shown that the thermal 
distortion in the beam splitter are complex due to the asymmetric beam
path~\cite{Wittel18}. While advanced thermal compensation systems
are being  investigated, it seems unlikely that alterative 1 would provide
sufficiently good contrast defect at the beamsplitter.
In the following we will therefore concentrate our investigation on
alternative 2.

It should be noted that 
a general optimisation
for detectors with other requirements in terms of power, bandwidth, 
cavity finesse, etc. can easily be undertaken using numerical models.
For example, the optical losses and their effect on the quantum-noise 
reduction is a key aspect of a more detailed design, see~\cite{Sabina2019}.

\section{Propagation of mirror position noise into the graviational
  wave channel}

In this section we investigate the noise coupling of the position of
the new optical elements, such as the sloshing cavity mirrors, into
the gravitational wave channel, i.e. the optical signal detected in
the dark port.

\begin{figure}[htb]
	\begin{center}
		\includegraphics[width=0.485\textwidth]{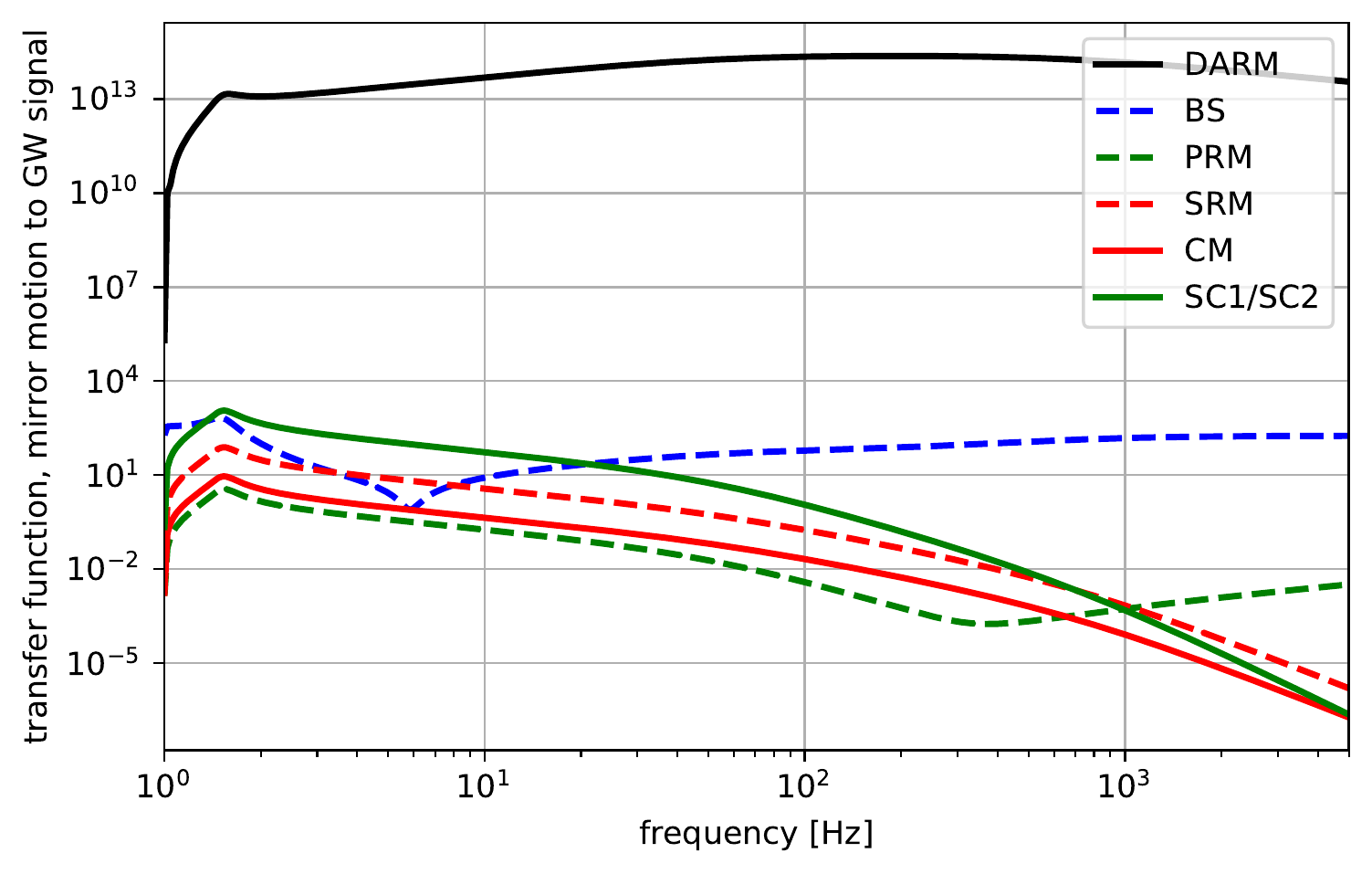}
	\end{center}
	\caption[]{Coupling of mirror displacement into the
    gravitational wave output channel. The plot shows the coupling of
    mirror motion of the components of the sloshing cavity (CM, SC1,
    SC2) and compares it to the coupling of the beamsplitter and
    recycling mirrors (BS, PRM, SRM). It can be seen that the noise
    coupling is of similar order of magnitude or smaller (at higher
    frequencies).} 
	\label{fig:noise_coupling}
\end{figure}

Figure~\ref{fig:noise_coupling} shows the simulated transfer functions for
mirror motions, comparing the motion of the sloshing cavity (CM, SC1,
SC2) to the coupling of the beamsplitter and
recycling mirrors (BS, PRM, SRM). For reference the coupling of a
differential end mirror motion (mimicking a gravitational wave signal)
is shown as well. 

It should be noted that the model does not
include any defects, such as unwanted asymmetries in the
interferometer arms.
Instead we have used a simple method to 
model the effects of such asymmetries on the noise couplings
by applying  a small offset to the differential arm
length of the interferometer. A so-called DC-offset is
used currently by LIGO in order to generate a local oscillator
field in the output port~\cite{AdvancedLIGO15}. 
The DC offset is a deliberate deviation
from the symmetric system set carefully such that it couples light 
into the output port with just enough power to dominate 
over any  other light leaking out of the interferometer due to unwanted
asymmetries.
Therefore the DC offset in the LIGO detectors provides a good 
indirect measure of the expected unwanted asymmetries we can expect.
In the next upgrade the LIGO detectors will be equipped with a homodyne
readout systems for which the DC offset is not required. 
We also  assume a homodyne readout for the interferometer
layouts presented here. 
An investigation using an interferometer model without defects but by
including the current amount of DC offset allows to mimic the
effect of interferometer defects in a simple and effective way.
Using this numerical model we evaluate the relative 
strength of the
coupling between different optics and provides an upper limit for the
absolute noise coupling.
This result shows that the coupling for the sloshing cavity
is at most of similar order of magnitude as the coupling of the
optics in the short Michelson interferometer (BS, SRM, PRM). Therefore
standard mirror suspension systems will be sufficient for the sloshing
cavity optics. Additionally, a sensing and control scheme for the sloshing cavity
mirror positions would have to achieve the same level of precision as
the control system for the Michelson optics. 

We suggest that the alternative configuration proposed in this
article should be further investigated for its behaviour in the
presence of defects as was done for the Sagnac configuration 
in \cite{danilishin15}. In addition, requirement on the mode matching
should be evaluated. Both are beyond the scope of this article.

\begin{acknowledgments}
The presented work was started during the very productive Speed Meter
Workshop in 2016, sponsored by the International Max-Planck Partnership
Scotland and hosted and organised by Stefan Hild. The authors would like
to thank Stefan Danilishin, Yanbei Chen, Paul Fulda and Conor Mow-Lowry for 
useful comments and discussions. A. Freise has been supported by the 
Science and Technology Facilities Council (STFC) and  
by a Royal Society Wolfson Fellowship which is jointly funded by the 
Royal Society and the Wolfson Foundation. Daniel Brown was supported
by the Australian Research Council grant CE170100004.

\end{acknowledgments}


\appendix

\section{Detector bandwidth}\label{sec:bandwidth}
In the following we recall characteristic parameters as
defined in~\cite{Purdue02}, based on the standard notation 
introduced in~\cite{Kimble02}. For current detectors the 
target shape of the quantum-noise-limited sensitivity curve 
is a wide 'bucket' around an optimal frequency around
100\,Hz. The width of the region with low quantum noise
is defined by the detector bandwidth, given by the
half-bandwidth  of the optical cavity storing the optical 
signal. In case of a Michelson with Fabry-Perot arm cavities 
this is given by the half-bandwidth of the arm cavities.
The half bandwidth, or pole frequency, of such a cavity can be computed as:
\begin{equation}
\gamma
= \frac{\rm c}{2 L}\arcsin\left(\frac{1-r_{\rm
      ITM}}{\sqrt{r_{\rm ITM}}}\right)
\end{equation}
with $T_{\rm ITM}$ the transmissivity  of the  input test mass (assuming
$T_{\rm ETM}\ll T_{\rm ITM}$), $c$ the speed of light and $L$ the
length of the arm cavities. In the cases discussed here, the
argument of the $\arcsin$ function is usually small so that 
we can approximate this as:
\begin{equation}
\gamma \approx 
 \frac{\rm c}{2L}\frac{1-r_{\rm
      ITM}}{\sqrt{r_{\rm ITM}}}
\end{equation}

In Advanced LIGO the half bandwidth is
given by the half bandwidth of the coupled cavity created by the 
signal recycling mirror (SRM) and the arm cavity mirrors (ITM, ETM).
We can understand the combination of SRM and ITM as 
one compound mirror $M_{\rm SRC}$. The 
power reflectivity and transmissivity of the cavity 
can be computed as:
\begin{equation}
R_{\rm SRC}= \frac{R_{\rm ITM} + R_{\rm SRM} -2r_{\rm ITM}r_{\rm
    SRM}\cos(\phi)}
{1+R_{\rm ITM}R_{\rm SRM}-2r_{\rm ITM}r_{\rm    SRM}\cos(\phi)}, 
\end{equation}
and
\begin{equation}
T_{\rm SRC}= \frac{T_{\rm ITM} T_{\rm SRM}}
{1+R_{\rm ITM}R_{\rm SRM}-2r_{\rm ITM}r_{\rm    SRM}\cos(\phi)}
\end{equation}
with $\phi$ the tuning of the cavity. The nominal Advanced LIGO design
assumes the SRC to be  tuned for resonant signal extraction with $\phi=0$
and we can simplify:
\begin{equation}
R_{\rm SRC}= \frac{(r_{\rm ITM} - r_{\rm SRM})^2}{(1-r_{\rm ITM}r_{\rm
    SRM})^2}, \,
T_{\rm SRC} = \frac{T_{\rm ITM}T_{\rm SRM}}{(1-r_{\rm ITM}r_{\rm SRM})^2}
\end{equation}

The half bandwidth of the detector, given by the half bandwidth of the cavity formed by the arm and the signal
recycling mirror, is then:
\begin{equation}
\gamma\approx \frac{c}{2 L} \frac{1-r_{\rm
      SRC}}{\sqrt{r_{\rm SRC}}}
\end{equation}

Purdue and Chen define two quantities to describe 
the sloshing speedmeter: the sloshing frequency $\Omega$
and the signal extraction rate $\delta$ as angular frequencies.
The sloshing frequency $\Omega$ indicates at which frequency 
the quantum-noise-limited sensitivity will be best and 
the extraction rate $\delta$ is equivalent to the bandwidth 
$\gamma$ in the Advanced LIGO layout. 

\section{Reflectivity of the sloshing cavity with the coupling mirror inside}
\label{app:sloshing_inside}

To compute the bandwidth of the interferometer in a 
sloshing configuration we need to know the reflectivity of the
sloshing cavity for a light field leaking out of the
interferometer. If the coupling mirror is located inside the
sloshing cavity we need to use a slightly different equation
than usual for the reflectivity of the cavity.

\begin{figure}[htb]
	\begin{center}
		\includegraphics[width=0.485\textwidth]{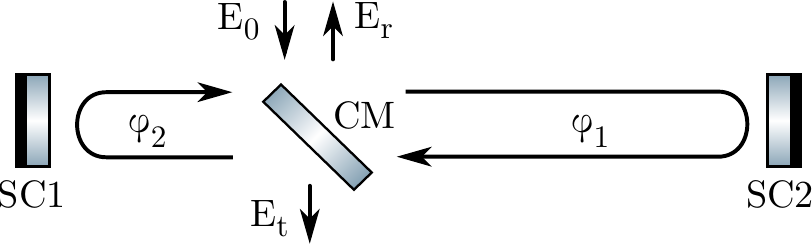}
	\end{center}
	\caption[]{Sketch to show the electric field coupling at the
    coupling mirror inside the sloshing cavity. $E_0$ is the incoming
    field from the main interferometer. $\varphi_1$ refers to the phase
    the light accumulates in the right half of the cavity, $\varphi_2$ is
for the left half.} 
	\label{fig:sloshing_reflectivity}
\end{figure}
A sketch of the relevant 
optical setup is shown in Figure~\ref{fig:sloshing_reflectivity}.
The reflectivity of the sloshing cavity is then defined as
$r_{\rm SC}= E_{\rm r}/E_0$ and the transmissivity as 
$t_{\rm SC}= E_{\rm t}/E_0$ respectively. 

Assuming almost perfectly reflecting mirrors SC1 and SC2, the
amplitude reflectivity and transmissivity can be computed from
the parameters of the coupling mirrors $R_{\rm CM}$, $T_{\rm CM}$ 
and the cavity tunings, given as $\varphi1$ and $\varphi2$. We use
the convention of applying a 90 degrees phase shift on transmission of
an optical element and none at reflection as described in
~\cite{Bond2017}.
We get:
\begin{equation}
r_{\rm SC}= \frac{R_{\rm CM} e^{- i \varphi_1} + R_{\rm CM} T_{\rm CM}e^{i \varphi_2}}{1+T_{\rm CM}^2}
\end{equation}
and
\begin{equation}
t_{\rm SC}= i\, t_{\rm CM}\,\frac{1+ e^{- i (\varphi_1+\varphi_2)}}{1+  T_{\rm CM}e^{i (\varphi_1+\varphi_2)}}
\end{equation}

If the sloshing cavity is on resonance for the carrier light frequency
the sum of the phases is known so that
\begin{equation}
\exp{\left(-i \left(\varphi_1 + \varphi_2\right)\right)} = -1
\end{equation}

The power coefficient are then independent of the differential tuning:
\begin{equation}
R_{\rm SC}= \frac{(1-T_{\rm CM})^2}{(1+T_{\rm CM})^2}
\end{equation}
\begin{equation}
T_{\rm SC}= \frac{4\,T_{\rm CM}}{(1+T_{\rm CM})^2}
\end{equation}

\bibliography{references}
\bibliographystyle{unsrt}

\end{document}